\documentclass[twocolumn,prl,amsmath,amssymb,superscriptaddress,showpacs,preprintnumbers]
              {revtex4}
\usepackage{graphics,bm,color,nicefrac,macros}

\begin{document}
\preprint{LMU-ASC 25/07}

\title{Coupling of transverse and longitudinal response in stiff
  polymers} \author{Benedikt Obermayer}
\email{obermayer@physik.lmu.de} \affiliation{Arnold Sommerfeld Center
  and Center for NanoScience, Ludwig-Maximilians-Universit\"at
  M\"unchen, Theresienstr. 37, 80333 M\"unchen, Germany} \author{Oskar
  Hallatschek} \email{ohallats@physics.harvard.edu} \affiliation{Lyman
  Laboratory of Physics, Harvard University, Cambridge, MA 02138, USA
} \date{\today}

\begin{abstract}
  The time-dependent transverse response of stiff inextensible
  polymers is well understood on the linear level, where transverse
  and longitudinal displacements evolve independently.  We show that
  for times beyond a characteristic time $\tf$, longitudinal friction
  considerably \emph{weakens} the response compared to the widely used
  linear response predictions.  The corresponding feedback mechanism
  is explained by scaling arguments and quantified by a systematic
  theory. Our scaling laws and exact solutions for the transverse
  response apply to cytoskeletal filaments as well as DNA under
  tension.  \pacs{61.41.+e, 87.15.La, 87.15.He, 98.75.Da}
\end{abstract}
\maketitle

In tracing back the viscoelasticity of the cell to properties of its
constituents, a detailed understanding of the mechanical response of
single cytoskeletal filaments is indispensable. Due to their large
bending stiffness, these filaments exhibit highly anisotropic
static~\cite{mackintosh-kaes-janmey:95} and
dynamic~\cite{gittes-mackintosh:98,morse:98,legoff-hallatschek-frey-amblard:02}
features, such as the anomalous $t^{\nicefrac{3}{4}}$-growth of
fluctuation amplitudes in the transverse
direction~\cite{amblard-etal:96,granek:97}, i.e., perpendicular to the
local tangent.  The related response to a localized transverse driving
force has so far been examined only by neglecting longitudinal degrees
of freedom~\cite{amblard-etal:96,wiggins-etal:98}, although these
polymers are virtually inextensible, and transverse and longitudinal
contour deformations therefore coupled. In this Letter we show that
longitudinal motion strongly affects the transverse response even for
weakly-bending filaments and leads to relevant nonlinearities beyond a
characteristic time $\tf$.

\begin{figure}[b]
  \centerline{\resizebox{.48\textwidth}{!}{\includegraphics{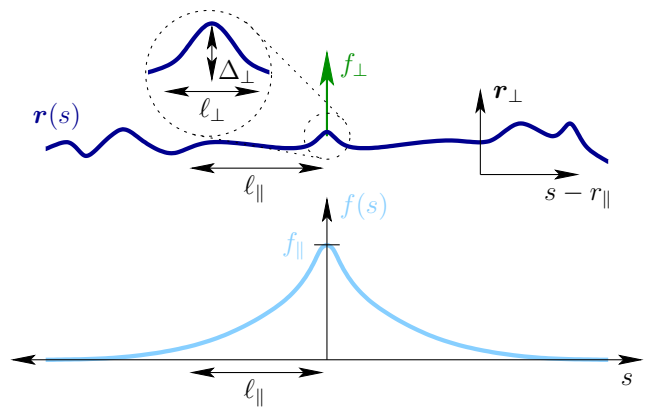}}}
  \caption{\label{fig:transverse}(Color online) A transverse point
    force $\fperp$ applied to the contour $\bvec r(s)$ (dark)
    translates, through the formation of a bulge of height
    $\Delta_\perp$ and width $\lperp$, into a longitudinal pulling
    force $\fpar$ acting on the polymer's tails.  This force induces
    backbone tension $f(s)$ (light) that penetrates the contour within
    a region of size $\lpar$ where thermal undulations are
    straightened.}
\end{figure}

The physical key factors controlling the transverse response may be
understood from Fig.~\ref{fig:transverse}, which shows a
weakly-bending polymer (bending undulations are exaggerated for
visualization) shortly after a transverse driving force $\fperp$ has
been applied in the bulk. In response to this force, the contour
develops a bulge.  Due to the backbone inextensibility, this bulge can
continue growing only by \emph{pulling in} contour length from the
filament's tails.  This effectively reduces the thermal roughness of
the
contour~\cite{seifert-wintz-nelson:96,brochard-buguin-degennes:99,everaers-juelicher-ajdari-maggs:99},
at a rate substantially limited by longitudinal solvent friction.  The
resulting coupling to the longitudinal response tends to \emph{slow
  down} the bulge growth. In order to describe this feedback
mechanism, we start with a scaling analysis and treat the simpler
\emph{athermal} case first. To connect to the biologically important
situations of prestressed actin networks~\cite{gardel-etal:06} and
prestretched DNA~\cite{bohbot_raviv-etal:04}, we then extend a recent
theory of tension dynamics~\cite{hallatschek-frey-kroy:05} to
calculate the nonlinear response for unstretched and prestretched
initial conditions.

\begin{table*}[t!]
  \begin{tabular}{|c|c|c|c|}
    \hline\hline
    \textbf{(a)} & $\lperp(t)$ & $\fpar(t)$ & $\Delta_\perp(t)$ \\
    \hline
    $t\ll \tf$ & $t^{\nicefrac{1}{4}}$ & 
    $\zeta L \fperp^2 t^{\nicefrac{1}{4}}$ & $\fperp t^{\nicefrac{3}{4}}$ \\
    $t\gg \tf$ & $[t\fpar(t)]^{\nicefrac{1}{2}}$ &
    $(\zeta L)^{\nicefrac{2}{5}} \fperp^{\nicefrac{4}{5}} t^{-\nicefrac{1}{5}}$ &
    $(\zeta L)^{-\nicefrac{1}{5}}(\fperp t)^{\nicefrac{3}{5}}$ \\
    \hline\hline
  \end{tabular}
  \hspace{.05cm}
  \begin{tabular}{|c|c|c|c|c|}
    \hline\hline
    \textbf{(b)} & $\lperp(t)$ & $\lpar(t)$ & $\fpar(t)$ & $\Delta_\perp(t)$ \\
    \hline
    $t\ll \tf$ & $t^{\nicefrac{1}{4}}$ & 
    $(\lp/\zeta)^{\nicefrac{1}{2}} t^{\nicefrac{1}{8}}$ &   
    $(\zeta\lp)^{\nicefrac{1}{2}}\fperp^2 t^{\nicefrac{3}{8}}$ & 
    $ \fperp t^{\nicefrac{3}{4}}$ \\
    $t\gg \tf$ &  $[t\fpar(t)]^{\nicefrac{1}{2}}$ & 
    $(\lp/\zeta)^{\nicefrac{1}{2}} [t\fpar(t)]^{\nicefrac{1}{4}}$ &
    $(\zeta\lp)^{\nicefrac{2}{9}} \fperp^{\nicefrac{8}{9}} t^{-\nicefrac{1}{9}}$ &
    $(\zeta\lp)^{-\nicefrac{1}{9}} (\fperp t)^{\nicefrac{5}{9}}$ \\
    \hline\hline
  \end{tabular}
  \caption{\label{tb:nonlinear-response} Summary of crossover scaling laws for an initially unstretched filament. The crossover time $\tf$ is implicitly defined through $\tf=\fpar^{-2}(\tf)$, $\fpar(t)$ is the induced longitudinal force, $\Delta_\perp(t)$ is the transverse response and $\ell_{\perp/\parallel}$ is the transverse/longitudinal correlation length~\cite{seifert-wintz-nelson:96,everaers-juelicher-ajdari-maggs:99,hallatschek-frey-kroy:05}. (a) Athermal case. $\fpar$ pulls in the filament's tails of length $L$. $\tf=(\gammaa\fperp)^{-2}$ with $\gammaa=(\zeta L)^{\nicefrac{2}{3}}\fperp^{\nicefrac{1}{3}}$. (b) Thermal case. The filament's tails have effective length $\lpar\ll L$. $\tf=(\gammat\fperp)^{-2}$ with $\gammat=(\zeta\lp)^{\nicefrac{2}{7}}\fperp^{\nicefrac{1}{7}}$.}
\end{table*}

Consider the overdamped dynamics of an initially straight stiff rod of
total length $L$. Suddenly applying a transverse pulling force
$\fperp$, for simplicity in the center of the rod, leads to the growth
of a bulge deformation. The generated friction in the transverse and
longitudinal direction needs to be balanced by corresponding driving
forces.  Viscous solvent friction is modeled via anisotropic friction
coefficients (per length) $\zeta_\perp$ and $\zeta_\parallel=\zeta
\zeta_\perp$ with $\zeta\approx \frac{1}{2}$~\cite{wiggins-etal:98}
for transverse and longitudinal motion, respectively. After a time
$t$, the resulting bulge has some characteristic height
$\Delta_\perp(t)$ and width $\lperp(t)$.  The transverse force
$\fperp$ balances the drag force $\zeta_\perp\lperp \Delta_\perp/t$
acting on a polymer section of length $\lperp$ moving transversely
with velocity $\Delta_\perp/t$ through the solvent; hence,
$\Delta_\perp\simeq \fperp t/(\zeta_\perp\lperp)$.  Naturally, the
contour length along the deformed rod section is larger than its
longitudinal extent $\lperp$.  Assuming a simple ``triangle'' geometry
as in the blow-up in Fig.~\ref{fig:transverse}, the difference is
roughly given by $\Delta_\perp^2/\lperp$. In order to provide this
\emph{stored} (or excess) length, the filament's tails are pulled in
by a longitudinal force $\fpar$.  The latter has to balance the
longitudinal friction that acts on the filament's tails of length $L$
moving longitudinally with a velocity given by the temporal change of
the excess contour length contained in the bulge.  Estimating
$\fpar\simeq\zeta_\parallel L \Delta_\perp^2/(\lperp t)$, we plug in
$\Delta_\perp$ from above and get $\fpar\simeq \zeta_\parallel
L\fperp^2 t/(\zeta_\perp^2\lperp^3)$.

The yet unknown time-dependent width $\lperp(t)$ of the bulge is
controlled by the relaxation spectrum of bending deformations. In the
weakly-bending limit, the transverse displacement field $\rperp(s,t)$
of an overdamped inextensible rod with bending stiffness $\kappa$
obeys~\cite{granek:97}
\begin{equation}\label{eq:athermal-transverse}
  \zeta_\perp\pd_t \rperp = -\kappa\,\rperp '''' + \fpar(t) \rperp''\;,
\end{equation}
in the presence of a longitudinal pulling force $\fpar(t)$. Primes
denote derivatives with respect to the arclength coordinate
$s\in[-\frac{L}{2},\frac{L}{2}]$. In the following, we set $\kappa$
and $\zeta_\perp$ to unity, such that time is a length$^4$ and force a
length$^{-2}$. From a simple scaling analysis of
Eq.~(\ref{eq:athermal-transverse}), $\rperp/t\simeq \rperp
(\lperp^{-4}+\fpar \lperp^{-2})$, we deduce the growing size
$\lperp(t)$ of a bending deformation (assuming $\lperp\ll L$).
Inserting appropriate formulas~\cite{hallatschek-frey-kroy:05} for
$\lperp(t)$ into the relations for $\Delta_\perp$ and $\fpar$ derived
before finally yields the selfconsistent scaling laws for $\fpar(t)$
and the nonlinear response $\Delta_\perp(t)$ summarized in
Table~\ref{tb:nonlinear-response}(a).  For short times the coupling
effect is irrelevant and $\Delta_\perp(t)$ is linear in $\fperp$.
However, this requires the small force $\fpar$ to pull in more and
more contour length from the tails and increases the longitudinal
friction to be balanced by $\fpar$.  At the crossover time $\tf$, this
force becomes large enough (typically, $\fpar\simeq \gammat
\fperp\gtrsim \fperp$) to feed back onto the transverse dynamics,
which is manifest in nonlinear dependencies~\footnote{Since the
  weakly-bending assumption still holds at $t=\tf$, longitudinal
  friction is the only relevant nonlinearity and higher-order terms
  $\propto \rperp^3$ in Eq.~\eqref{eq:eom-mspt-T} are negligible.}  on
$\fperp$.  In particular, it considerably slows down the bulge growth,
which in turn requires $\fpar$ to pull in contour length at a slower
rate and eventually makes it decrease.
  
\begin{figure}
  \centerline{\resizebox{.5\textwidth}{!}{\includegraphics{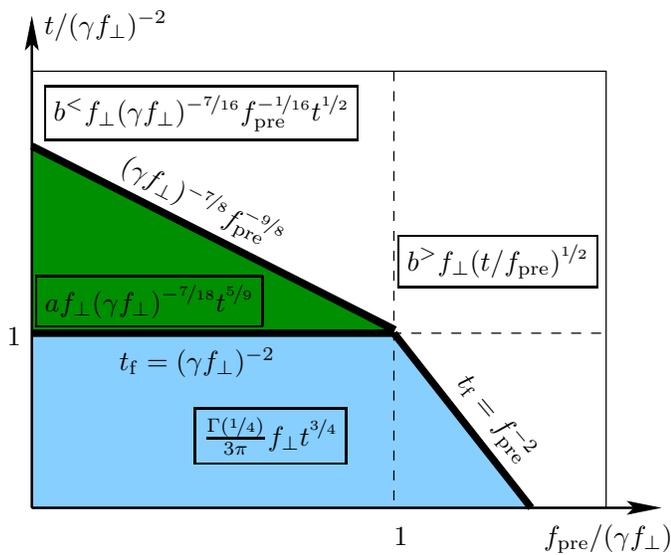}}}
  \caption{\label{fig:prestretched}(Color online) Regimes of
    intermediate asymptotics (separated by thick black lines) for the
    nonlinear response $\Delta_\perp(t)$ (boxed formulas); time
    $t/(\gammat\fperp)^{-2}$ vs.\ force ratio $\fpre/(\gammat\fperp)$
    (log-log scale).  The universal initial regime~\cite{granek:97}
    (light shaded) is followed by a quasi-static regime (white) with
    different force scaling for asymptotically small $(<)$ and large
    $(>)$~\cite{granek:97} force ratio; in these limits, the
    respective prefactors are
    $b^{<}\sim[8(1+\sqrt{2})^2/\pi^3]^{\nicefrac{1}{8}}$ and
    $b^{>}\sim\pi^{-\nicefrac{1}{2}}$. An intermediate taut-string
    regime (dark shaded) emerges for very small force ratio. The
    prefactor is $a=[3(2+\sqrt{2})/\pi^2]^{\nicefrac{2}{9}}$ if
    $\fpre=0$.}
\end{figure}

The essential difference for nonzero temperatures is the presence of
thermal contour undulations, see Fig.~\ref{fig:transverse}, which are
correlated over the persistence length $\lp=(\kb T)^{-1}$, and
straightened out by the longitudinal force $\fpar$.  Still
counteracted by longitudinal friction, this happens first only within
a small but growing region of size $\lpar(t)$ (see
Refs.~\cite{seifert-wintz-nelson:96,morse:98,
  brochard-buguin-degennes:99,everaers-juelicher-ajdari-maggs:99,
  hallatschek-frey-kroy:05}). Correspondingly, the force $\fpar(t)$
from above has to be generalized to a tension field $f(s,t)$, which
decays over the length scale $\lpar(t)$. Crossover scaling laws for
$\lpar(t)$, shown in Table~\ref{tb:nonlinear-response}(b), were
derived for constant external force in
Ref.~\cite{hallatschek-frey-kroy:05} and can be generalized to
(weakly) time-dependent ``external'' forces such as $\fpar(t)$.  The
thermal problem is essentially analogous to the athermal case for late
times $t > \tlpar$ where $\tlpar$ is defined via $\lpar(\tlpar)=L$.
However, if the region $\lpar(t)$, where the contour straightens, does
not yet extend to the filament's ends ($\lpar\ll L$, or $t \ll
\tlpar$), the ``thermal'' rod has only an effective time-dependent
length of $\lpar(t)$.  Hence, scaling laws for the nonlinear response
are then obtained simply by replacing $L\to\lpar$ in
Table~\ref{tb:nonlinear-response}(a), which gives the results
summarized in Table~\ref{tb:nonlinear-response}(b).  These apply to
initially unstretched filaments while the general case of prestretched
initial conditions is discussed below and summarized in
Fig.~\ref{fig:prestretched}.  Naturally, the replacement $L\to\lpar$
affects only the long-time scaling of the nonlinear response
$\Delta_\perp(t)$ -- on short times $t\ll\tf$, the transverse dynamics
evolves undisturbed by the longitudinal one.  We expect the
anomalously slow long-time response to be observable in many
biological situations. In aqueous solution, we roughly estimate a
crossover time $\tf\approx
10^{-2}\,\mathrm{s}/\fperp[\pN]^{\nicefrac{8}{3}}$ for typical
microtubules with $L\approx 10\,\mathrm{\mu
  m}$~\cite{pampaloni-etal:06} (representing the athermal case).
Under thermal conditions, where the ``interesting'' time window is
between $\tf$ and $\tlpar$, we get $\tf\approx
10^{-3}\,\mathrm{s}/\fperp[\pN]^{\nicefrac{16}{7}}$ and $\tlpar\approx
0.2\,\mathrm{s}/\fperp[\pN]$ for (unstretched) actin filaments of
about $20\,\mathrm{\mu m}$
length~\cite{legoff-hallatschek-frey-amblard:02}, which implies that
the actin response to myosin motors becomes nonlinear on time scales
comparable to the duration of a single power
stroke~\cite{tyska-warshaw:02}. Filaments in actin networks (mesh size
$\xi\approx \frac{1}{10}L\approx 0.5\mum$) under stresses of about
$1\,\mathrm{Pa}$~\cite{gardel-etal:06} are usually so short that
$\tf\gg\tlpar\approx 10^{-4}\,\mathrm{s}$, but the coupling
nonlinearity should be observable in the viscoelastic
response~\cite{morse:98}. Finally, $\tf\approx
10^{-5}\,\mathrm{s}/\fperp[\pN]^{\nicefrac{16}{7}}$ and $\tlpar\approx
0.05\,\mathrm{s}/(\fperp[\pN]\fpre[\pN]^{\nicefrac{5}{8}})$ for DNA
($L\approx 20\,\mathrm{\mu m}$~\cite{bohbot_raviv-etal:04})
prestretched with $\fpre\ll\fperp$.

In order to support and quantify the scaling picture developed above,
we proceed with a systematic approach similar to
Ref.~\cite{hallatschek-frey-kroy:05} based on the length scale
separation $\lpar(t)\gg\lperp(t)$. As long as the dynamics induced by
the transverse force is not influenced by end effects ($\lpar\ll L$),
we consider a semi-infinite arc\-length interval, $s\in[0,\infty)$,
and represent the transverse force as a \emph{boundary condition} at
$s=0$. In the wormlike chain Hamiltonian,
$\mathcal{H}=\frac{1}{2}\int\!\td s\, [\bvec r''^2 + f\bvec r'^2]$,
the tension $f(s,t)$ enforces the local inextensibility constraint
$\bvec r'^2(s,t)=1$.  Parametrizing the contour $\bvec
r(s,t)=(\rperp,s-\rpar)^T$ by its transverse and longitudinal
displacements from a straight line (see Fig.~\ref{fig:transverse}),
the weakly-bending limit of small contour gradients
$\rperp'^2=\mathcal{O}(\eps)\ll 1$ is realized for very stiff polymers
(
$\eps\equiv L/\lp$), alternatively for semiflexible filaments strongly
prestretched with a force $\fpre$ (
$\eps\equiv \fpre^{-\nicefrac{1}{2}}/\lp$).

The conformational dynamics in solution follows from a balance of
elastic and tensile forces $-\delta\mathcal{H}/\delta \bvec r$,
thermal noise $\bvec\xi$, and anisotropic friction $[\bvec r'\bvec
r'+\zeta(1-\bvec r'\bvec r')]\pd_t\bvec r$~\cite{wiggins-etal:98}.
Within the weakly-bending limit, transverse and longitudinal
fluctuations have strongly different correlation lengths:
$\lperp/\lpar=\mathcal{O}(\eps^{\nicefrac{1}{2}})$; cf.\
Table~\ref{tb:nonlinear-response}(b).  An adiabatic approximation
(justified via a multiple scale analysis) exploits this scale
separation. The resulting equations of
motion~\cite{hallatschek-frey-kroy:05} are written in terms of
formally independent rapidly and slowly varying arc\-length parameters
$s$ and $\bar s \eps^{\nicefrac{1}{2}}$, respectively:
\begin{subequations}\label{eq:eom-mspt}
  \begin{align}\label{eq:eom-mspt-T}
    \pd_t \rperp &= - \pd_s^4\rperp + \bar f \pd_s^2\rperp +
    \bvec\xi_\perp +\bfperp \delta(s)\Theta(t), \\
    \label{eq:eom-mspt-L}
    \pd_{\bar s}^2 \bar f &= -\zeta \ave{\pd_t\overline\varrho}.
  \end{align}
\end{subequations}
Eq.~\eqref{eq:eom-mspt-T} gives the small-scale dynamics of the
transverse displacements $\rperp(s,t)$ for \emph{locally constant}
tension $f\equiv\bar f(\bar s,t)$, cf.\
Eq.~\eqref{eq:athermal-transverse}. Using a Cosine transform with
respect to $s$, it is readily solved by the response function
\begin{equation}\label{eq:chi-perp}
  \chi_\perp (q;t,t') = \e^{-q^2[q^2 (t-t') + \int_{t'}^t\!
    \td \tau\bar f(\bar s,\tau)]} 
  \Theta(t-t').
\end{equation}
Eq.~\eqref{eq:eom-mspt-L} describes the coarse-grained tension
variations on the large scale $\bar s \eps^{\nicefrac{1}{2}}$: it
relates curvature in the tension to (average) changes in stored length
density $\ave{\overline\varrho}(\bar
s,t)\equiv\ave{\overline{\frac{1}{2}\rperp'^2}}(\bar s,t)$. Averaged
both thermally and spatially (on the small scale $s$),
$\ave{\overline\varrho}$ inherits its remaining $\bar s$-dependence
from the tension $\bar f$ in Eq.~\eqref{eq:chi-perp}:
\begin{equation}
  \label{eq:varrho}
  \ave{\overline\varrho}= \ave{\frac{1}{2}\left[\int_{0}^\infty\! 
      \frac{\td q}{\pi}\int_{-\infty}^t\!\!\!\td t'q \chi_\perp(q;t,t')
      \bvec\xi_\perp(q,t')\right]^2}.
\end{equation}
Reintroducing a single unique arclength variable, $\bar s\equiv s$,
Eqs.~\eqref{eq:eom-mspt-L} and \eqref{eq:varrho} result in a nonlinear
partial integro-differential equation (PIDE) for $\bar f(s,t)$, that
was analyzed in Ref.~\cite{hallatschek-frey-kroy:05} for explicitly
prescribed boundary conditions.  In the present case, however, the
boundary condition at $ s=0$ has to be determined implicitly. The
polymer's inextensibility 
requires that the bulge be created using stored length from the tails.
To formalize this condition, we demand at any time a vanishing average
longitudinal velocity $\ave{\pd_t\rpar}$ at the origin where the force
is applied, and also at infinity.  Inextensibility
($\rpar'=\frac{1}{2}\rperp'^2+\mathcal{O}(\eps^2) \approx \varrho$)
gives $0=\int_0^\infty\!\!\td s\,\langle\pd_t\rpar'\rangle=
\int_0^\infty\!\!\td s\ave{\pd_t\varrho}$. With $\pd_{ s}\bar
f\rvert_{s\to\infty}=0$ and Eq.~\eqref{eq:eom-mspt-L}, this constraint
implies
\begin{equation}
  \label{eq:bc}
  \pd_{s} \bar f\rvert_{s=0} =-\zeta\int_0^\infty\!\!\td
  s\,\pd_t\ave{\varrho-\overline\varrho}.
\end{equation} 
The difference $ \ave{\varrho-\overline\varrho}$ represents the excess
length stored in the bulge on the small length scale $\lperp$.
Consequently, it did not contribute to Eq.~\eqref{eq:varrho} which was
spatially coarse-grained on intermediate scales $\lperp\ll l \ll
\lpar$. It can be obtained, though, from the right hand side of
Eq.~\eqref{eq:varrho} upon replacing $\bvec\xi_\perp \to -\bfperp\sin
q s\,\Theta(t)$. Evaluating the $s$-integral in Eq.~(\ref{eq:bc}) to
leading order yields our central analytical result: a boundary
condition for the tension that quantifies the feedback between
``bulge'' and ``tail'' dynamics:
\begin{equation}
  \label{eq:pide-bc}
  \pd_s \bar f\rvert_{s=0} = 
  -\frac{\zeta\fperp^2}{4}\int_0^\infty\!\!
  \frac{\td q}{\pi}\pd_t
  \left[\int_0^t\!\td t'\,q\chi_\perp(q;t,t')\rvert_{s=0}\right]^2.
\end{equation}

\begin{figure}
  \centerline{\resizebox{.5\textwidth}{!}{
      \includegraphics{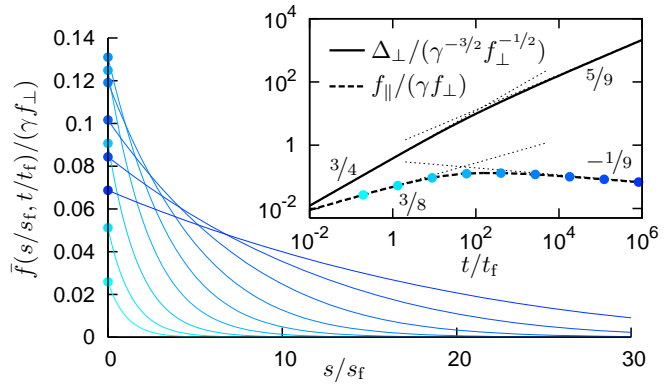}}}
  \caption{\label{fig:transverse-numerical} (Color online) Numerical
    solutions $\bar f(s,t)$ to
    Eqs.~(\ref{eq:eom-mspt-L},\ref{eq:varrho},\ref{eq:pide-bc}) for
    $\fpre=0$, time is increasing from light to dark color.  Inset:
    log-log plot of the effective longitudinal force $\fpar(t)=\bar
    f(0,t)$ (circles/dashed), and of the nonlinear response
    $\Delta_\perp(t)$ from Eq.~\eqref{eq:nonlinear-response} (solid).
    Dotted lines indicate the asymptotes of
    Table~\ref{tb:nonlinear-response}(b).  The crossover scales are
    $\tf=(\gammat\fperp)^{-2}$ and
    $\sf=(\lp/\zeta)^{\nicefrac{1}{2}}(\gammat\fperp)^{-\nicefrac{1}{4}}$,
    with
    $\gammat=(\lp\zeta)^{\nicefrac{2}{7}}\fperp^{\nicefrac{1}{7}}$.}
\end{figure}

In terms of the response function $\chi_\perp(q;t,t')$ of
Eq.~\eqref{eq:chi-perp}, the average displacement $\Delta_\perp(t)$
induced by the transverse force (i.e., the nonlinear response) reads
\begin{equation}\label{eq:nonlinear-response}
  \Delta_\perp(t)=
  \fperp \int_0^\infty\!\!\frac{\td q}{\pi}\int_0^t\!\!\td t'
  \chi_\perp(q;t,t')\rvert_{s=0},
\end{equation}
which is evaluated at $s=0$ after the tension profiles $\bar f(s,t)$
are computed from
Eqs.~(\ref{eq:eom-mspt-L},\ref{eq:varrho},\ref{eq:pide-bc}). To this
end, we introduce two-variable scaling
forms~\cite{hallatschek-frey-kroy:05} that remove any parameter
dependence: $\bar f(s,t)=\gammat\fperp\, \varphi(s/\sf,t/\tf)$, with
the crossover scales $\tf$ and $\sf$ and $\gammat$ as in
Fig.~\ref{fig:transverse-numerical}. Numerical solutions are obtained
by mapping the PIDE onto a system of nonlinear
equations~\cite{obermayer-hallatschek-frey-kroy:07}. Selected tension
profiles are displayed in Fig.~\ref{fig:transverse-numerical} and
describe one half of the filament with $\fperp$ being applied at the
origin. Our analytical approach is based on reducing the scaling forms
$\varphi(s/\sf,t/\tf)$ to one-variable scaling functions
$\varphi\sim(t/\tf)^\alpha\hat\varphi(s/\lpar(t))$ with $\lpar(t)=\sf
(t/\tf)^z$ in the asymptotic limits of short and long times.  In the
latter limit $t\gg \tf$, we recover either the taut-string
approximation of Ref.~\cite{seifert-wintz-nelson:96} and may neglect
bending and thermal forces, or the quasi-static approximation of
Ref.~\cite{brochard-buguin-degennes:99}, which lets us treat the
tension as locally equilibrated. Which approximation is valid depends
quite strongly on the prestretching force $\fpre$ through the ratio
$\fpre/(\gammat \fperp)$, similar to the related scenario of
longitudinal stretching forces applied to prestretched
filaments~\cite{obermayer-hallatschek-frey-kroy:07}.  The resulting
intermediate asymptotic scaling laws for $\Delta_\perp(t)$ are
summarized in Fig.~\ref{fig:prestretched}, including analytical
prefactors.  For a given ratio $\fpre/(\gammat \fperp)$, the evolution
of $\Delta_\perp(t)$ corresponds to a vertical path through
Fig.~\ref{fig:prestretched}.  The exact solutions quickly converge to
these asymptotes, as shown in the inset of
Fig.~\ref{fig:transverse-numerical} for the limiting case $\fpre=0$.

In summary, we argue that the coupling between transverse and
longitudinal response affects not only single polymers, but also
single crosslinks, crosslinked networks, and tensegrity
structures~\cite{morse:98,everaers-juelicher-ajdari-maggs:99,gardel-etal:06,ingber:03a}.
For completeness, we note that our self-consistent approach both for
the heuristic ``bulge'' idea as well as for the systematic derivation
of Eq.~\eqref{eq:pide-bc} applies only to the
\emph{non}linear~\cite{hallatschek-frey-kroy:05} response on
sufficiently small times $t\ll \tlpar,\tc$.  At $\tlpar$, end effects
become important, and at $\tc$, the weakly-bending assumption breaks
down: the contour gradients become large when
$\Delta_\perp\simeq\lperp$.  We find that $\tc \gtrsim\tlpar$ for
initially weakly-bending filaments (as those in the above discussed
situations)~\footnote{$\tc\simeq (\lp/L)^9\tlpar$ if
  $\fperp\gg\lp^{-2}$ and $\fpre\lesssim\lp^{-2}$; otherwise $\tc \gg
  \tlpar$. For $\fperp\gg\lp^{\nicefrac{3}{2}}/L^{\nicefrac{7}{2}}$
  ($\fpre\gg\lp^2/L^4$), $\tlpar$ falls into the taut-string
  (quasi-static) regime (cf.\ Fig.~\ref{fig:prestretched}).}. Our
analysis of the generic coupling mechanism is not constrained by the
details of the relaxation regime $t\gg\tlpar$ (which is similar to the
athermal case).

\acknowledgments We thank K.~Kroy, E.~Frey, T.~Munk, and C.~Heussinger
for helpful discussions. O.H. acknowledges financial support by the
German Academic Exchange Program (DAAD) and by the Deutsche
Forschungsgemeinschaft (DFG) through grant no.~Ha 5163/1. B.O. is
supported by the DFG through SFB 486, by the German Excellence
Initiative via the program ``Nanosystems Initiative Munich (NIM)'',
and through BayEFG.

\end{document}